\documentclass[useAMS]{mn2e}
\usepackage{graphicx}
\title[Ultra Long GRBs]{Are Ultra Long Gamma Ray Bursts powered by black holes spinning down?}
\author[Nathanail et al.]{Antonios Nathanail$^{1,2}$\thanks{E-mail:
antonionitoni@hotmail.com}
and Ioannis Contopoulos$^{1}$ \\
$^{1}$Research Center for Astronomy and Applied Mathematics,
Academy of Athens, Athens 11527, Greece\\
$^{2}$Section of Astrophysics, Astronomy and Mechanics,
Department of Physics, University of Athens, \\
Panepistimiopolis Zografos, Athens 15783, Greece}	

\def\gsim{\mathrel{\raise.5ex\hbox{$>$}\mkern-14mu
             \lower0.6ex\hbox{$\sim$}}}
\def\lsim{\mathrel{\raise.3ex\hbox{$<$}\mkern-14mu
             \lower0.6ex\hbox{$\sim$}}}

\begin{document}

\date{Accepted . Received ; in original form }

\pagerange{\pageref{firstpage}--\pageref{lastpage}} \pubyear{2015}

\maketitle

\label{firstpage}

\begin{abstract}

Gamma-ray bursts (GRBs) are violent explosions, coming from
cosmological distances.  They are detected in gamma-rays (also
X-rays, UV, optical, radio) almost every day, and have typical
durations of  a  few seconds to a  few minutes. Some  GRBs have
been reported with  extraordinary durations of $10^4$~sec, the
so-called Ultra Long GRBs. It has been debated whether these form
a new distinct class of events or whether they are similar to long
GRBs. According to Blandford \& Znajek (1977), the spin energy of
a rotating black hole can be extracted electromagnetically, should
the hole be endowed with a magnetic field supported by  electric
currents  in a surrounding  disk. We argue that this can be the
case for the central engines of GRBs and we show that the duration
of the burst depends on the magnetic flux accumulated on the event
horizon of the black hole. We thus estimate the surface magnetic
field of a possible progenitor star, and we conclude that an Ultra
Long GRB may originate from a progenitor star with a relatively
low  magnetic field.

\end{abstract}


\section{Introduction}

Gamma-ray bursts (hereafter GRBs),  are cosmic flashes of
gamma-rays, and consist some of the most energetic events ever
detected, with luminosities  exceeding $10^{50}$~erg/sec.
Multi-wavelength observations of these enigmatic events allowed us
to go deeper into the underlying physics (Gerhels \& Ramirez-Ruiz
2009), but still a lot more work is needed to form a complete
picture of them. Even more interesting are  events that lasted a
lot longer than the usual. Ultra Long GRBs have durations of $
10^4$~sec, when typical durations for long GRBs are a few seconds
to a few minutes in the observer frame. Several Ultra Long
Gamma-ray bursts have been reported (Levan et al. 2014, Evans et
al. 2014, Gendre et al.
2013)\footnote{http://www.astro.caltech.edu/grbox/grbox.php
(online GRB catalog comments the very long ones).}.

A great amount of theoretical work has been invested in order to
understand what is the central engine and the emission mechanism
of GRBs. In recent years the question of the central engine has
been put aside, while research  focuses on the emission region,
the emission mechanisms and the effort to understand all the
characteristics of the light curves and the spectra of the bursts
(recent review Kumar \& Zhang 2014). The idea of a black hole
powering the burst is widely discussed for the so-called
long-duration GRBs.

For the usual long GRBs, the progenitor of the burst  is thought
to be a Wolf-Rayet star. This is rather hard to verify since, by
the time we detect the burst, the progenitor is not in its
previous form. For Ultra Long GRBs, the idea of a blue supergiant
(Gendre et al. 2013) is also put into play to explain the long
lived duration, which is proposed to be due to the accretion of a
massive hydrogen envelope (Nakauchi et al. 2013).
In both models though, the resulting collapse to a compact
object (black hole or neutron star) is inevitable.

The extraction of rotational energy from a black hole through the
Blandford \& Znajek mechanism (1977) has been  studied in great
depth. The black hole spins down and looses angular momentum,
giving  off electromagnetic (Poynting) energy that is somehow
dissipated into high energy radiation. An important question is
what is the timescale for the spin down. In the presence of strong
magnetic fields this process will last for thousands of seconds,
whereas for ultra strong magnetic field strengths it could last
for only a fraction of a second (Lee et al. 2000, Contopoulos et
al. 2014). As we will now see, the duration of the spin down is
inversely proportional to the square of the magnetic flux
accumulated on the black hole event horizon.

 \begin{figure}
\includegraphics[width=84mm]{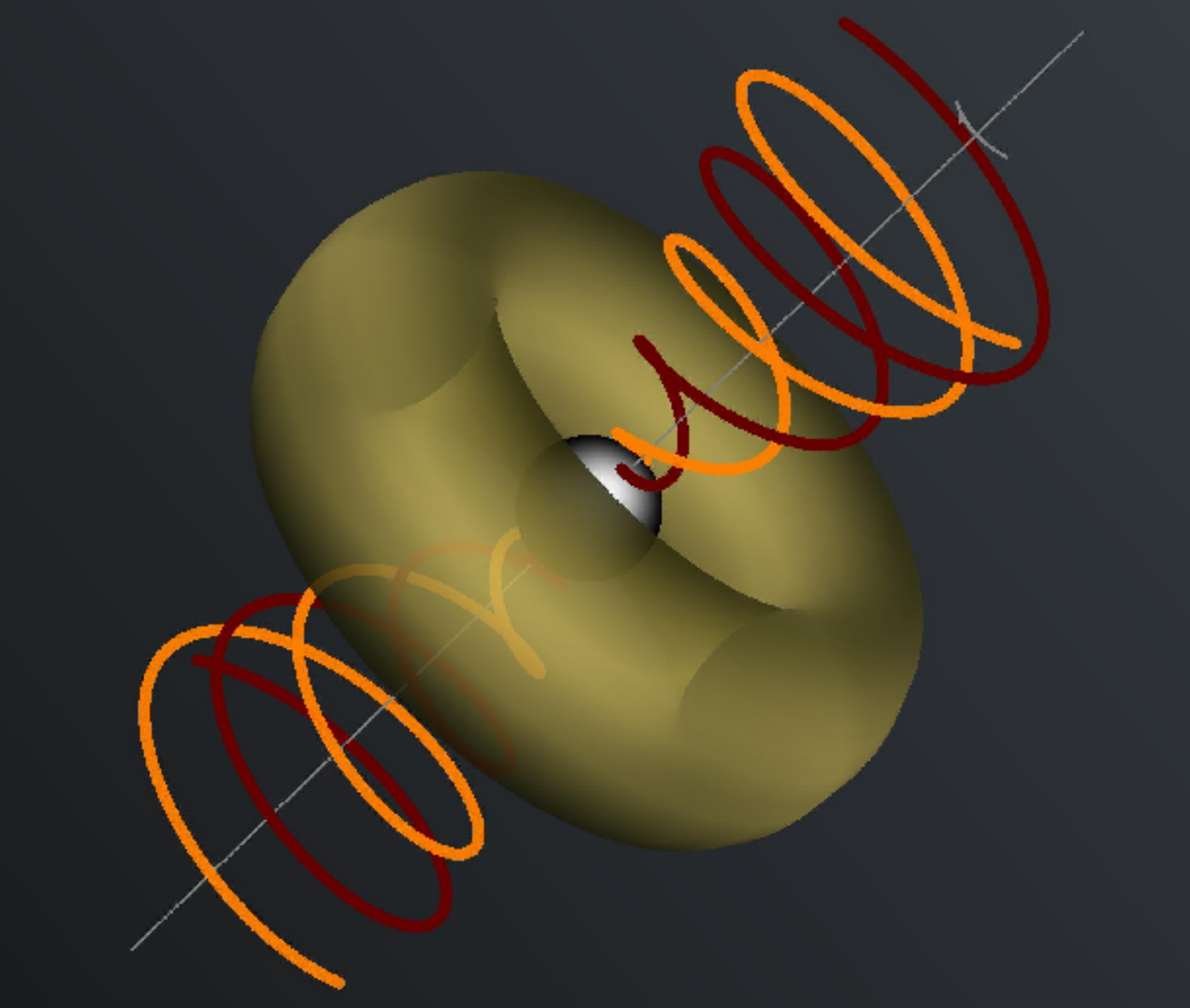}
\caption{The structure of the magnetosphere close to the event
horizon of a rotating black hole. Magnetic field lines (depicted
in  dark red and orange) based on the solutions of Nathanail \&
Contopoulos~(2014). A massive torus of material (transparent)
holds the magnetic flux on to the event horizon.}
 \label{BHTRYLAST}
 \end{figure}

\section{Black Hole Spin Down}

Let us consider a supermassive progenitor star whose core
collapses and forms a rotating black hole. It is natural for the
star to be magnetized. Highly conducting matter from the interior
of the star will drive the advection of magnetic flux during the
collapse. The material that is going to collapse will be strongly
magnetized, and therefore its core will pass through a spinning
magnetized neutron star stage. A certain amount of magnetic flux
$\Psi_m$ is then going to cross the horizon. An equatorial  thick
disk (torus) will form around the black hole due to the rotational
collapse. A black hole cannot hold its own magnetic field, but the
material from the thick disk will act as a barrier that will hold
the magnetic flux initially advected.

As long as this is the case, the black hole will lose
rotational/reducible energy at a rate
\begin{equation}
\dot{E} \approx -\frac{1}{6\pi^2 c}\Psi_m^2\Omega^2\ ,
\label{EdotIa}
\end{equation}
and will thus spin down very dramatically (Blandford \& Znajek
1977 for low spin parameters; Tchekhovskoy et al. 2010,
Contopoulos et al. 2013, Nathanail \& Contopoulos 2014 for
maximally rotating black holes). $\Omega$ is the angular velocity
of the black hole horizon. In principle, this procedure can
extract almost all the available/reducible energy
\begin{equation}
E_{\rm rot}=M c^2 - M_{\rm irr} c^2\ ,
\end{equation}
where $M_{\rm irr}$ is the irreducible black hole
mass\footnote{$M_{\rm irr}=\sqrt{A_H c^4 /16 \pi G^2}~$ where
$A_H$ is the surface area of the black hole.} (Christodoulou \&
Ruffini 1971, Misner, Thorne \& Wheeler~1973). The reader can
check that the rotational energy of a $10M_{\odot}$ initially
maximally rotating black hole is $E_{\rm rot}\approx 5\times
10^{54}$~erg, which is a rather extreme value for the total energy
released in a GRB explosion (Komissarov, personal communication).
However, if the black hole is e.g. rotating at $10\%$ of maximum,
then $E_{\rm rot}\approx 2\times 10^{52}$~erg which is much more
reasonable for a GRB. It is clear that if we change the mass and
the spin of the black hole, the energy it can give off spans more
than three orders of magnitude.

In what follows, we will assume that the newly formed
black hole is slowly rotating. Under that approximation,
$M\approx$~const. and
\begin{equation}
E_{\rm rot} \approx \frac{1}{8}Mc^2\left(\frac{\Omega}{\Omega_{\rm
max}}\right)^2\ , \label{Erot}
\end{equation}
where $\Omega_{\rm max}\equiv c^3/2{\cal G} M$ is the angular
velocity of a maximally rotating black hole, and ${\cal G}$ is the
gravitational constant. The black hole will, therefore, spin down
as
\begin{equation}
\dot{E} = \frac{{\cal G}^2M^3}{2c^4}\frac{{\rm d}(\Omega^2)}{{\rm
d}t}\ .\label{EdotIb}
\end{equation}
Equating eqs.~(\ref{EdotIa}) and (\ref{EdotIb}) and solving for
$\Omega=\Omega(t)$, we obtain
\begin{equation}
\dot{E} \propto e^{-t/t_{BZ}} \label{Eapprox}
\end{equation}
where
\begin{equation}
t_{BZ}\equiv \frac{3 c^5}{16{\cal G}^2 B^2 M}=50
\left(\frac{B}{10^{15}\ \mbox{G}}\right)^{-2} \left(\frac{M}{10
M_\odot}\right)^{-1}\mbox{sec}\ \label{tBZ}
\end{equation}
is the timescale for the spinning down procedure. We have defined
here a typical value for the accumulated black hole magnetic field
required by the Blandford-Znajek mechanism,
\begin{equation}
B =\frac{\Psi_m c^4}{4\pi {\cal G}^2 M^2}\ .
\end{equation}
$B$ can reach very high values during the core-collapse of a
massive star, and for $B\sim 10^{15}$~G, the black hole spins down
in a few hundred seconds. It is interesting to notice here that
eq.~(5) cannot distinguish between a black hole and a
neutron star/magnetar with field lines that are held open by the
surrounding material. Therefore, in principle, we cannot claim
that exponential decay GRBs is definite proof for the presence of
a central black hole (it is the same as pulsar spin down with
breaking index equal to unity). It is, however, a strong
suggestion since, if a cavity around the central object expands
sufficiently for a light cylinder to appear in it, a magnetar will
follow the standard power law pulsar spindown which will be
different from the exponential one that we are investigating in
this work.

We will now argue that it is very reasonable for this
magnetic field to be held in place by a massive disk/torus of
material of mass M$_{\rm d}$ and angular momentum per unit mass
$l_{\rm d}$. A crude calculation of the force balance between the
outward electromagnetic force, gravity and rotation yields
\begin{equation}
\frac{B^2}{r}r^3 \sim \frac{G M M_{\rm d}}{r^2} - \frac{M_{\rm d}
l_{\rm d}^2}{r^3} \label{Bdisk}
\end{equation}
where $r$ is the radius and approximate height of the torus. If
the disk is rotationally supported, eq.~(8) does not
allow for any extra magnetic field to be held in its interior.
This could be the case for a progenitor star with relatively fast
rotation. If, on the other hand, the progenitor star is not
rotating as fast, a slowly rotating black hole may form at the
center (as we argued above), while the rest of the left over
stellar material may not have enough angular momentum to form a
centrifugally supported disk around it (Woosley \& Heger 2006). In
that case, it is natural to imagine that the equilibrium described
by eq.~(8) is reached. One can easily check that, in
order to support a magnetic field strength of $B\sim 10^{15}$~G
for very small values of $l_{\rm d}$, a torus of size $r\sim
2{\cal G}M/c^2$ and mass $M_d\sim 10^{-5}~M_{\odot}$ around a
$10M_{\odot}$ black hole is all that is needed. For higher values
of $l_{\rm d}$ one needs a higher torus mass to hold the same
value of the magnetic field. Notice that we are not presently
considering the stability of this configuration against e.g.
Rayleigh-Taylor instability (Contopoulos \& Papadopoulos 2012). We
just assume that it survives for the duration of the black hole
spin down that we propose we are observing in a GRB.

Our present discussion has an indirect implication for the
so-called `efficiency' of the Blandford-Znajek mechanism usually
defined as
\begin{equation}
\eta=\frac{\dot{E}}{\dot{M}_{\rm d}c^2} \label{eta}
\end{equation}
(e.g. Tchekhovskoy, Narayan \& McKinney 2011). Here, $\dot{M}_{\rm
d}$ is the accretion rate in the disk of material around the
central black hole, and numerical simulations have shown that
$\eta$ can even surpass unity (e.g. Tchekhovskoy \&
McKinney~2012)! This indirectly reveals the obvious fact that
accretion is not sufficient to drive the system, and an extra
source of energy (black hole rotation) is tapped. We have argued
above that, in general, $B$ is unrelated to $\dot{M}_{\rm d}$,
essentially decoupling the Blanford-Znajek mechanism from the
accretion process. In that case, $\eta$ can in principle reach
infinitely large values in GRBs. On the other hand, in AGNs and
X-ray binaries, it is customary to assume that $B$ is held in
place not by gravity but by accretion, which is itself linearly
related to the mass of the central black hole. Only in this
scenario does it make sense to define and calculate an efficiency
for the Blanford-Znajek mechanism.

 \begin{figure}
\includegraphics[width=84mm]{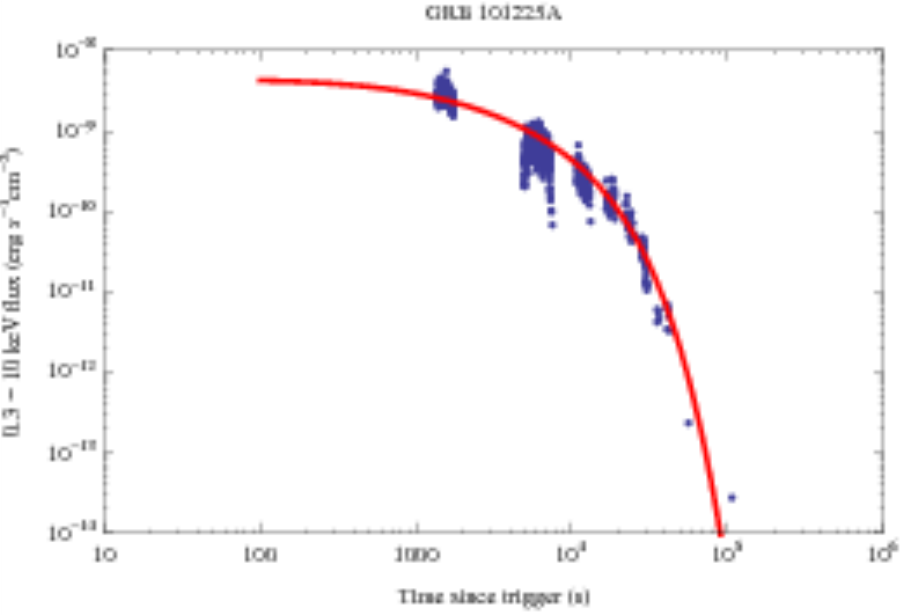}
\caption{Ultra Long $GRB ~101225A$ light curve.  Log-Log plot. The
red curve is the theoretical exponential black hole spin down.
Energy flux  at $0.3~-~10$~keV.}
 \label{ULGRB3}
 \end{figure}

Many effects can modify the black hole electromagnetic spin down,
making it difficult to discern its activation and evolution. GRB
events  may be `contaminated' by extra events that possibly take
place during the spin down. One such possibility is that large
enough mass infalls may result in sudden black hole spin ups, with
subsequent different electromagnetic spin downs. Such secondary
events will begin from a different peak of the light curve, thus
it is difficult to estimate the new spin parameter that the spin
down will start from. Also, if the massive disk is dispersed
faster than the duration of the spin down, the accumulated
magnetic flux $\Psi_m$ will not be conserved, and the spin down
evolution will not be exponential. Notice that the electromagnetic
interaction
 with the torus formed  around the
black hole may result in an extra spin down that may too be
linked to GRBs (van Putten et al. 2009).

The aforementioned calculation describes how the system looses
energy in the form of electromagnetic Poynting flux, and does not
account for the detailed radiation emission mechanisms. In
Nathanail \& Contopoulos (2014) we studied the structure of the
magnetic field in the vicinity of the black hole (Fig.~1). We
found that a generic feature of black hole magnetopsheres is a
poloidal electric current sheet that originates on the horizon and
extends along the last open magnetic field line that crosses the
horizon at the equator. Obviously $\gamma \gamma$~opacity will
produce a great amount of electron-positron pairs near the
horizon, and will prevent high energy radiation from escaping.
Further out, as the current sheet is naturally collimated along
the axis of rotation by the surrounding stellar material, it
reaches the emission region where the opacity has fallen enough
for radiation to escape. Our observational and theoretical
experience from pulsars suggests that high energy radiation is
expected to originate from reconnection processes that result in
particle acceleration along the magnetospheric current sheet
(Lyubarsky et al. 2001, Kalapotharakos et al. 2012, Sironi et al.
2014). We do expect a similar process in this generic  black hole
magnetospheric current sheet.


\section[]{Ultra Long GRBs}

According to the theoretical implications in the previous section,
we searched the new population of Ultra Long GRBs  (Levan et al.
2014) and looked for signs of overall black hole spin down as
described mathematically by eq.~(5). Many effects can
modify the spin down in the violent environment of a GRB (e.g.
mass infalls that result in sudden black hole spin ups), making it
difficult to discern its activation and evolution. Notice
that we focus on black hole spin down since other alternatives
such as neutrino annihilation (Leng \& Giannios 2014) fail to
explain the observed energetics of Ultra Long GRBs.

$GRB ~101225A$ (the `Christmas-day burst'): It is located at
redshift $z=0.847$. {\em Swift} saw this source from the very
beginning of its activity, after one {\em Swift} orbit ($90 ~$
min) the source was still active suggesting a very long duration.
This burst is reported as an Ultra Long GRB with
$T_{90}$\footnote{$T_{90}$ is the time since BAT triggered  till
the time that $90 \%$ of the counts are detected. Therefore, it is
an instrument dependent measure. Other measurements have been
proposed to define the time of central engine activity (Zhang et
al. 2014). } more than $1377$~sec (Campana et al. 2011, Thone et
al. 2011). The {\em Swift} XRT light curve are  taken from the
Swift /XRT team website (Evans et al. 2009) at the UK Swift
Science Data Centre (UKSSDC). Our theoretical fit for the black
hole spin down (eq.~5) yields the red curve in figure
3.

$GRB~ 111209A$: As BAT (the instrument of the {\em Swift}
satellite that detects in the energy band $15 - 350$~keV and
triggers for possible GRBs) was not looking at the part of the sky
from which this GRB arrived, it was triggered  after $150$~sec of
the actual time that this event was detectable (shown by the
Konus-Wind instrument Hoversten et al. 2011). It is located  at
redshift of z=0.677  (Vreeswijk et al. 2011). This burst is
reported as an Ultra Long GRB with $T_{90}$ around $800$~sec. The
total duration of the GRB activity is somewhat different from
$T_{90}$, for this burst lasts for more than $20000$~sec  (Gendre
et al. 2013). The {\em Swift} XRT light curves are  taken from the
Swift /XRT team website (Evans et al. 2009) at the UK Swift
Science Data Centre (UKSSDC). We fit the light curve with our
theoretical function (eq.~5). As  data are sparse, we
do not expect the theoretical curve to pass from all points but
rather to show that they  are following this prescription and that
the system is loosing energy according to this loss rate. As we
discussed before, we do not intent to fit all the  flares (such as
the one at around $400$~sec and another around $2000$~sec) of the
light curve with a multi-segment power law, but rather  to follow
the decrease in the energy flux with one function which has a
physical interpretation. In figures~3 \& 4 we show the light curve
of $GRB ~111209$A  (energy flux  at $0.3~-~10$~keV) together with
our theoretical exponential fit. The plot in figure ~4 is linear
in time, and shows clearly that around $50000$~sec the energy flux
stops decreasing and enters a plateau phase where the flux seems
constant for the next $50000$~sec. At this point it seems that
either the black hole continues to spin down but other mechanisms
cover its further evolution, either we are entering an afterglow
activity possibly dominated by external shocks.

The third candidate (namely $GRB ~121027$A) of the new population
of Ultra Long GRBs (Levan et al. 2014) has a more complicated
structure. The signs of exponential black hole spin down are still
evident, but the light curve, especially the big X-ray flare at
$10^3$~sec, needs further analysis (Wu et al. 2013). That is why
we decided not to include it here.

 \begin{figure}
\includegraphics[width=84mm]{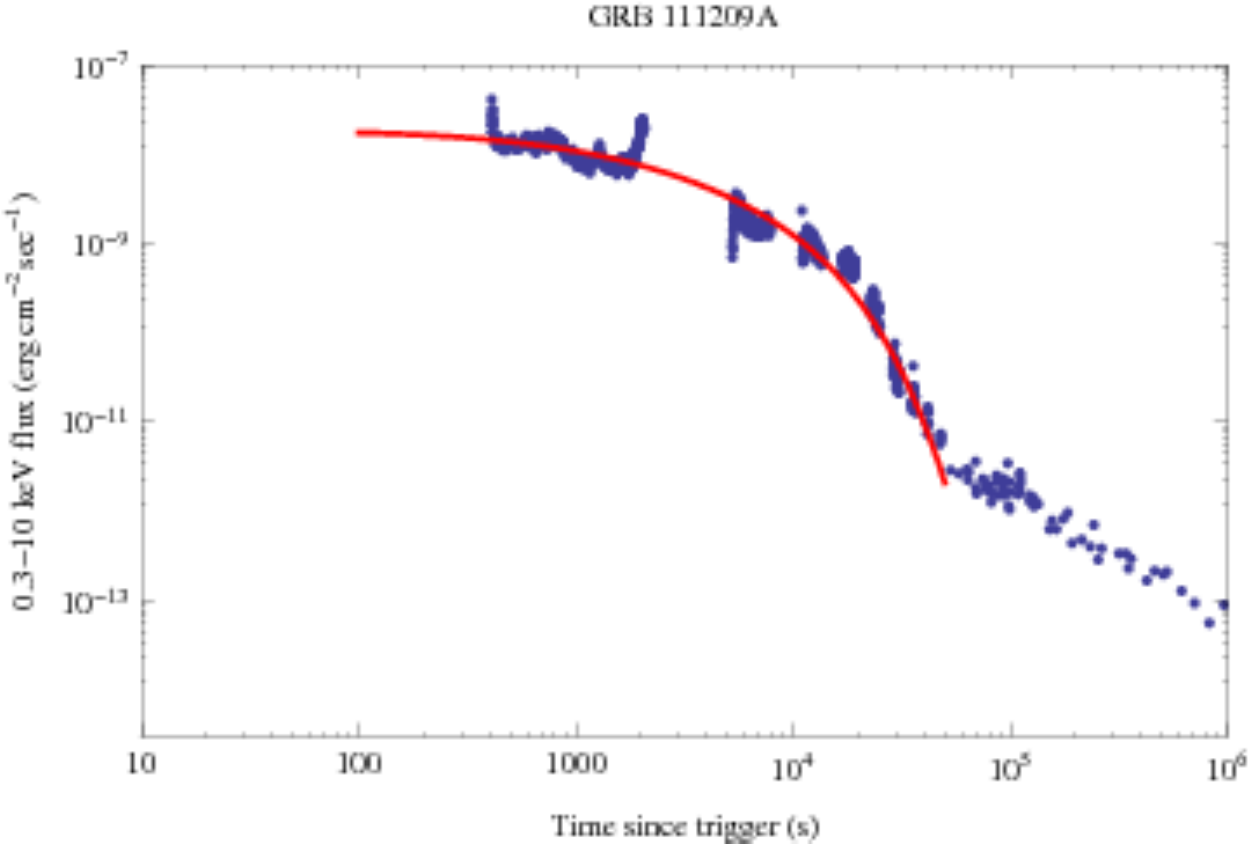}
\caption{Ultra Long $GRB ~111209A$ light curve.  Log-Log plot. The
red curve is the theoretical exponential black hole spin down.
Energy flux  at $0.3~-~10$~keV.}
 \label{ULGRB2}
 \end{figure}

 \begin{figure}
\includegraphics[width=84mm]{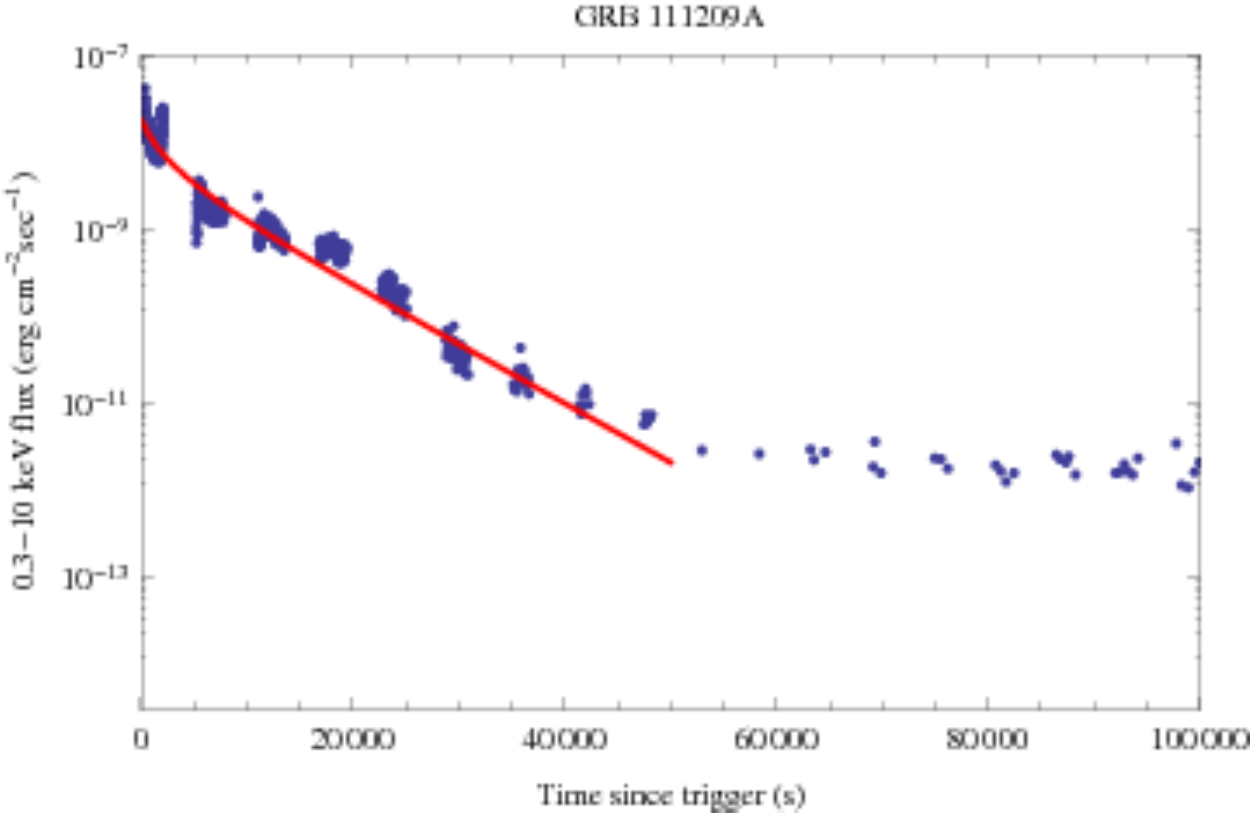}
 \caption{Ultra Long $GRB ~111209A$ light curve.  Log-Linear plot.
 Around
$50000$~sec the flux becomes constant. We believe that we can
follow the spin down of the black hole till that time. The red
curve is the theoretical exponential black hole spin down. Energy
flux  at $0.3~-~10$~keV.}
 \label{ULGRB1}
 \end{figure}

Our fit of the light curve allows us  to estimate $t_{BZ}$, which
is a really important physical parameter. Knowing $t_{BZ}$, we can
estimate the strength of the magnetic field in the vicinity of the
black hole. This can give us an estimate for the actual strength
of the magnetic field on the surface of the star before the
collapse and compare our findings with Wolf-Rayet polarization
measurements and magnetic field estimations. By fitting
eq.~(5) to the light curve and correcting for
cosmological redshift, we find that $t_{BZ} \approx 3400$~sec for
$GRB ~101225A$ and $t_{BZ} \approx 3600$~sec for $GRB ~111209A$.
We emphasize once again that $t_{BZ}$ is different from $T_{90}$
(which is detector dependent) and gives the timescale of the spin
down. The time that this exponential decay ends (when the energy
flux enters a plateau phase) can be related with a one to one
correspondence with $t_{\rm burst}$ defined in Zhang et al.
(2014). As $t_{BZ}$ is almost the same for the two bursts  we will
continue
 with a single  discussion covering both cases.

\bigskip
\begin{tabular}{|l|c|}
{\em GRB} & {\em $t_{BZ}$} (sec) \\
 \hline
 $101225A$  & $3400~(\pm  270)~$  \\
 \hline
   $111209A$ & $3600~(\pm  260)~$ \\
\hline
\end{tabular}
\bigskip

In order to continue our estimations, we assume that the black
holes formed after the core collapse have masses of $10
M_{\odot}$. This is a natural choice if the progenitor star is
$25~-~40~M_{\odot}$ (Heger et al. 2003).
Applying these values to eq.~(6), we find that the
estimated magnetic field on the event horizon is
\begin{equation}
B \approx 10^{14}~ G.
\end{equation}
While collapsing, the conducting matter of the stellar interior
brings this flux to the event horizon. Due to magnetic flux
conservation we have
\begin{equation}
B r^2 = B_{\star} r_{\star}^2\ , \label{Bstar}
\end{equation}
where $B_{\star}$ and $r_{\star}$ is the surface magnetic field
and the radius of the star respectively. A typical radius for  a
Wolf-Rayet star is $10^{12}$~cm (Crowther 2007), in which case
eq.~(11) yields $B_{\star}\sim 10^2$~G. Notice that this
estimate does not take into account possible dynamo magnetic
field amplification under the cataclysmic conditions in the
collapsing environment, as discussed in the literature
(Obergaulinger et al. 2009). If we assume an extra three orders
 dynamo field amplification our estimate of the surface
magnetic field may be as low as $B_{\star}\sim 0.1$~G.

According to a study of circular polarization and a search for
magnetic fields in Wolf-Rayet stars, the most probable field
strength in the observable part of its stellar wind is likely on
the order of  $10$ to $100$~G. Magnetic fields values of $~22 ~ -
~ 128$~G have been reported in the stellar winds of Wolf-Rayet
stars (de la Chevroti‘ere et al. 2014). These magnetic field
estimations are obtained from measurements of emission lines in
the  stellar winds and not on the stellar surface. At visible
wavelengths, the stellar surface of Wolf-Rayet stars is hidden by
a dense nebula. The corresponding surface value of the magnetic
field must be much higher than the observed estimated values. In
order to compare with observations we need to estimate  the
magnetic field in the stellar wind where the field is stretched
into a monopole configuration and drops as $1/r^2$ with distance.
Under this assumption,  the magnetic field in  the stellar wind,
ten stellar radii from the surface,  would be on the order $0.001$
to $1$~G. This  calculation leads us to believe that these bursts
may very well be  coming from a progenitor Wolf-Rayet star of a
really low  magnetic field strength and this may be the reason of
its ultra long duration.

The above estimates were obtained with the physical image of a
Wolf-Rayet star discussed extensively in the GRB literature
(Woosley \& Bloom 2006). Even if our model for the progenitor star
changes, our proposition that the duration of these bursts depends
on the magnetic field  will still hold. The idea that magnetic
flux is the principal parameter that sets the luminosity of a GRB
is discussed also in Tchekhovskoy \& Giannios (2015), although in
their case the central engine  turns-off when the steep decline
stage starts.

\section[]{Conclusion}

Our results suggest that the duration of the central engine's
activity depends on the magnetic flux accumulated   on the event
horizon of the newly formed black hole after the core collapse of
a suppermassive star. This in turn depends on the surface magnetic
field of the progenitor star. Based on these ideas we suggest that
Ultra Long GRBs lie in the same class together with the usual long
GRBs, and their extraordinary duration is due to the low surface
magnetic field of the progenitor star.

\section*{Acknowledgements}

We acknowledge discussions with Serguei Komissarov, Jonathan
McKinney, and the referee Dimitris Giannios. This work made use of
data supplied by the UK Swift Science Data Centre at the
University of Leicester, and was supported by the General
Secretariat for Research and Technology of Greece and the European
Social Fund in the framework of Action `Excellence'.

{}

\end{document}